\newcommand{\haeop}{Ly$\alpha$\ }
\newcommand{\haeMpc}{\,{\rm Mpc}}
\newcommand{\haeyr}{\,{\rm yr}}
\newcommand{\haekms}{{\rm km\,s^{-1}}}
\newcommand{\haeMsol}{\,{M_\odot}}
\newcommand{\haesr}{\,{\rm sr}}
\newcommand{\haes}{\,{\rm s}}
\newcommand{\haecm}{\,{\rm cm}}
\newcommand{\haekeV}{\,{\rm keV}}
\newcommand{\haenW}{\,{\rm nW}}

\documentclass[runningheads]{cl2emult}

\usepackage{makeidx}  
\usepackage{graphicx} 
\usepackage{subeqnar} 
\usepackage{multicol} 
\usepackage{cropmark} 
\usepackage{eso}      
\makeindex            

\begin{document}
\title*{The formation and evolution of supermassive 
black holes and their host galaxies
\footnote{Invited talks presented at the  ESO workshop on 
``Black Holes in Binaries and Galactic Nuclei'' in Honour of Prof. 
R. Giacconi, Garching 1999}}
\toctitle{The formation and evolution of supermassive 
black holes \protect \newline and their host galaxies 
}
%
%
\titlerunning{Supermassive black holes
and their host galaxies}
%
\author{Martin G. Haehnelt and Guinevere Kauffmann\inst{\null}}
\authorrunning{Martin G. Haehnelt and Guinevere Kauffmann}
%
%
\institute{Max-Planck-Institut f\"ur Astrophysik,
Karl-Schwarzschild-Str. 1, 85740 Garching, Germany}

\maketitle    

\begin{abstract}
We discuss constraints on the assembly history of 
supermassive black holes from the observed remnant black 
holes in nearby galaxies and from the emission caused by accretion 
onto these black holes. We also summarize the results of a specific 
model for the evolution of galaxies and their central black holes
which traces their hierachical build-up in CDM-like cosmogonies. 
The model assumes (i) that black  holes, ellipticals and starburts 
form during major mergers of galaxies (ii) that the gas fraction in 
galaxies decreases with decreasing redshift (iii) that the optical 
bright phase of a QSO lasts for about $10^{7}$ years. The model 
succesfully reproduces the evolution of cold gas as traced 
by damped \haeop systems, the evolution of optically bright QSOs, 
the remnant black hole mass distribution  and the host galaxy  
luminosities of QSOs.

\end{abstract}

\section{Introduction}

The evidence for the existence of supermassive black holes 
has been steadily  
increasing over the last years. For the two most convincing 
cases our own galactic centre and NGC4258 
\cite{hae:gen97,hae:wat94,hae:miy95}, the evidence is now 
beyond resonable doubt. The evidence that most nearby 
galaxies contain supermassive black holes is also compelling. 
Early suggestions of a linear relation between the black hole 
mass and the bulge mass have been corroborated by larger samples
\cite{hae:kor95,hae:mag98}. It is generally believed that we observe a
significant fraction if not all the material falling into supermassive
black holes by the radiation emitted by active galactic nuclei. A supermassive 
black hole therefore seems  to ``know'' in  which galaxy it will end up
at the present day. This and the fact that supermassive black holes 
contain as much as 0.2 to 0.6 percent of the baryonic mass  
of the galaxy \cite{hae:mag98,hae:vdm98}, suggests that the 
the formation of stars in the bulges of galaxies and the assembly  
of supermassive black holes at their centre are closely linked 
\cite{hae:ric98,hae:chr99}. On the other hand, there is strong evidence that 
structures in the  Universe form hierarchically, i.e. larger structures
build up by merging of smaller structures. This is a 
generic feature of a wide class of structure formation scenarios, 
the so called cold dark matter (CDM) cosmogonies. Both galaxies 
and AGN activivity have been succesfully modelled
within such hierarchical cosmogonies 
\cite{hae:er88,hae:hr93,hae:kau96,hae:hai98,hae:kch98,hae:bau98,hae:spf99}.
Here we first review observational constraints on the accretion
history of supermassive  black holes and 
discuss some clues for the formation mechanism.  We then summarize 
the results of  a specific model that describes the joint evolution of 
galaxies and  supermassive  black holes 
within a hierarchical cosmogony (see Kauffmann \& Haehnelt 1998 
for more details \cite{hae:kha99}). 
We assume here $\Omega_{\rm mat} = 0.3$,  $\Omega_{_{\Lambda}} = 0.7$, 
$h =0.65$ and $\sigma_{8} =1$.

\section{The formation and evolution of supermassive black holes}

\subsection{Black hole mass densities}

We can get some information on the assembly  history of supermassive 
black holes in nearby galaxies by comparing the mass density of 
remnant black holes to that required to produce the radiation 
emitted during the accretion process 
\cite{hae:sal99}. The mass density in 
remnant black holes can be inferred from the total mass density in 
bulges and the mass ratio of black hole mass 
to bulge mass,
\[
\rho_{_{\rm BH}} =
7.2\times 10^{5} 
\left (\frac{M_{\rm bh}/M_{\rm bulge}}{0.002}\right )\,
\left (\frac{\Omega_{\rm bulge}}{0.003}\right ) \,
M_{\odot}\,\haeMpc^{-3}. \nonumber  
\]
The total mass density in bulges has been estimated 
by Fugukita Hogan and Peebles  to be  
$ 0.001h^{-1} \le \Omega_{\rm bulge} \le 0.003h^{-1}$ \cite{hae:fhp98}.
The normalization of the bulge  to black hole mass correlation 
is still a matter of debate. Magorrian et al claim a value 
of 0.6 percent while van der Marel argues that a value 
of 0.2 percent is more realistic \cite{hae:mag98,hae:vdm98}. 
We will adopt the latter value in the rest of the paper.
  
The integrated emission by optically bright QSOs 
due  to accretion onto supermassive black holes can also 
be used to infer the correponding mass density in 
supermassive black holes if an  efficiency for the 
transformation of accreted rest mass into optical $f_{_{\rm B}} \epsilon$
light is assumed \cite{hae:sol82,hae:cht92}, 
\[
\rho_{_{\rm Opt}}=
1.4\times 10^{5}\,
\left (\frac{f_{_{\rm B}}\,\epsilon}{0.01}\right )^{-1}\,
M_{\odot}\,\haeMpc^{-3}.\nonumber
\]
Here $\epsilon$ is the overall efficiency of transforming accreted rest
mass energy into radiation and $f_{\rm B}$ is the fraction emitted 
in the B-band. 
Similarly we can estimate the black hole mass density which results 
from the emission of hard X-rays \cite{hae:dim97}, 
\[
\rho_{_{\rm X-ray}} =
3.8\times 10^{5}\, \left (\frac{f_{_{\rm X-ray}}\,
\epsilon}{0.01}\right )^{-1}\,
M_{\odot}\,\haeMpc^{-3},\nonumber
\]
where we have assumed a total hard X-ray flux of 
$140 \haekeV \haes^{-1} \haecm^{-2} \haesr ^{-1}$. 
The sources  producing the hard X-ray background are 
generally assumed to be a  class of AGN 
different to optically bright QSOs that has not yet 
been identified. We adopt an effective emission redshift 
$z_{\rm em} = 1.5$ for these unidentified sources.  

Part of the IR-background should also be produced by AGN, 
although it has been argued that their contribution should  
not exceed 30 percent \cite{hae:alm99}. If 30 
percent of the IR background 
were indeed emitted by AGN \cite{hae:pug96}, then  
\[
\rho_{_{\rm IR}} =
7.5\times 10^{5}\,
\left (\frac{f_{_{\rm IR}}\,\epsilon}{0.1}\right )^{-1}\,
M_{\odot}\,\haeMpc^{-3},\nonumber
\]
where we have assumed a total IR flux of  
$15 \haenW \haes^{-1} \haecm^{-2} \haesr ^{-1}$ 
and again $z_{\rm em} = 1.5$. There has been 
some debate if the mass density inferred from the optical 
emission is large enough to explain the mass density in remnant
black holes alone. Even for the low value of black hole 
to bulge mass ratio of 0.2 percent adopted here, 
there is still a discrepancy of a factor of about five. 
This suggests (i) that there is either a significant contribution to the
black hole mass density by accretion other than that traced by optical
bright QSOs or (ii) that the efficiency for producing optical light
during the accretion is lower than usually assumed or (iii) 
that the black hole to bulge mass ratio is  lower than 0.2 percenct 
\cite{hae:phi97,hae:hnr98}. 
The possible additional accretion may well explain 
the hard X-ray backgound and part of the infrared background. 
It may or may not trace the 
evolution of optical bright quasars. 

\begin{figure}[b]
\centering
\includegraphics[width=1.0\textwidth]{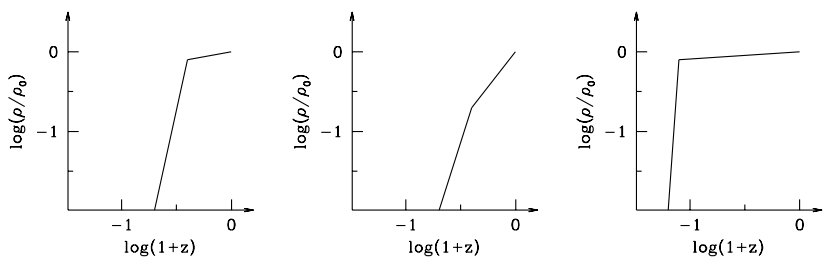}
\caption[]{A sketch of possible assembly histories of supermassive
black holes in nearby galaxies. Plotted is the  overall black hole mass 
density relative to its present-day value. For furter description 
see text.}
\end{figure}

\subsection{Possible assembly  histories of supermassive black holes}

There  is still a rather wide range of possible assembly 
histories for the supermassive black holes in nearby galaxies
(see Haehnelt, Natarajan \& Rees for a more detailed 
discussion \cite{hae:hnr98}).In Fig. 1 we sketch three possible 
options out of this range. In the {\em left panel} most of the mass 
is assembled in supermassive black holes during the epoch  of optically 
bright QSOs around $z\sim 2.5$. In the {\em middle panel}   
only 20 \% of the mass is assembled in supermassive black holes 
during the epoch  of optically bright QSO. The rest is accreted 
at low redshift, possibly in the form of hot gas in an advection dominated 
accretion flow. In the {\em right panel} most of the mass is assembled 
into small supermassive black holes  at very high redshift. 
Present-day supermassive black holes form predominatly by merging 
of smaller black holes. Accretion of gas during the epoch of bright 
QSOs or at an later epoch does not change the mass density much.

\subsection{Clues for the formation mechanism of the 
typical supermassive black hole} 

A variety of physical mechanisms for assembling 
mass into supermassive black holes have been suggested 
(see Rees 1984 \cite{hae:ree84} and Rees these proceedings 
for a  review),
\begin{itemize}
\item{the dynamical evolution of a dense cluster of stellar objects,}  
\item{the build-up of a supermassive black hole by merging 
of smaller  black holes,}  
\item{the viscous evolution and/or merger-driven 
collapse of  a self-gravitating gaseous object.} 
\end{itemize}

All of these processes certainly  occur and can lead to the formation
of supermassive black holes. The last option seems, however, the most 
attractive way of explaining the observed black hole bulge mass relation.
The main reason is the high formation efficiency inferred from the 
large black hole to bulge mass ratios. It is hard to see 
how as much as one percent of all available cold gas could end up
in a supermassive black hole of $10^{9}\haeMsol$  if an 
intermediate state of 
a dense stellar cluster is involved. Initially relaxation 
times in such a cluster would be long and a considerable fraction 
of stars would   evaporate before the cluster becomes dense enough 
to evolve rapidly \cite{hae:qui90}. It is also problematic 
to build up  supermassive black holes  predominantly by merging 
of smaller black holes that have formed well 
before the epoch of optically bright 
QSOs. This would require large black hole  formation efficiencies 
in shallow potential 
wells with $v_{\rm c} < 100 \haekms$. It seems  more plausible 
that black holes should form  less efficiently in 
smaller potential wells due to the feedback 
of the energy relased both by accretion onto the (forming)
supermassive black hole and due to supernovae.  
It is also unclear whether supermassive black holes in galaxies merge 
efficiently or whether sling-shot ejection plays a role 
\cite{hae:beg80,hae:mer99}.  We therefore consider the  assembly 
history in the right panel of Fig. 1 
to be rather improbable.

In  the next section   we will discuss a specific model for the evolution
of galaxies and their supermassive black holes within a hierarchical 
cosmogony. This model assumes that the optical bright QSOs do trace the 
accretion history of supermassive black holes well
(as in the left panel of Fig.1), but it could easily be altered to
accomodate other accretion modes.

\begin{figure}[t]
\centering
\includegraphics[width=1.0\textwidth]{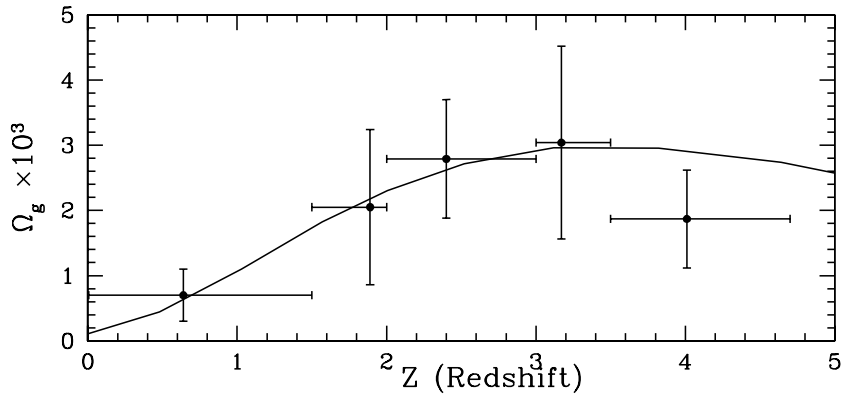}
\caption[]{
The cosmological mass density in cold gas in galaxies as a function of
redshift. The data is taken from 
Storrie-Lombardi \cite{hae:sto96}.}
\end{figure}

\section{Modelling the assembly of supermassive black holes 
and their host galaxies}

\subsection{Merging galaxies, starbursts and AGN}

\begin{figure}
\centering
\includegraphics[width=1.0\textwidth]{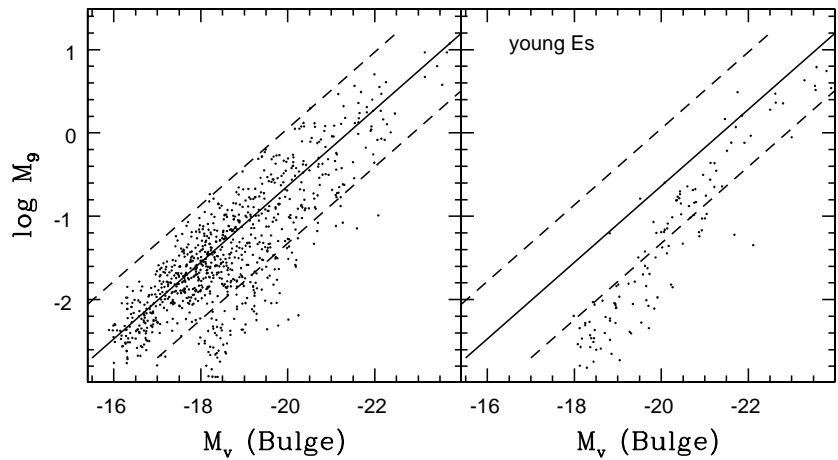}
\caption[]{ The correlation between the 
logarithm of the mass of the central black hole expressed in units 
of $10^9 M_{\odot}$ and the absolute V-band magnitude of the bulge. 
The dots are an absolute V-band magnitude limited sample of bulges 
in our model. The thick solid line is the $M_V({\rm bulge})$ vs. 
$M_{{\rm BH}}$ relation obtained by Magorrian et al \cite{hae:mag98} 
for nearby  normal galaxies. The dashed lines give an indication of the 
1$\sigma$ scatter in the observations. The right panels is the 
prediction for young ellipticals.}
\end{figure}

In CDM-like cosmogonies, galaxies build up by hierarchical merging. 
The formation  and evolution of galaxies in such cosmologies 
has been studied extensively using Monte-Carlo realizations 
of the hierarchical build-up of galaxies which include
simple prescriptions  to describe gas cooling, star formation, 
supernova feedback and merging rates of galaxies. These models 
reproduce many observed properties of galaxies both at low 
and at high redshifts \cite{hae:kau96,hae:kch98,hae:bau98,hae:spf99}. 
In the models, the quiescent accretion of gas from the halo results 
in the formation of a disk.
If two galaxies of comparable mass merge, a spheroid is formed
and the remaining gas undergoes a starburst. 
We assume here that such major mergers are also responsible for the 
growth and fuelling of black holes in galactic nuclei. If two galaxies of 
comparable mass merge, the central  black holes of the progenitors  
coalesce and a few percent of the  gas in the merger remnant is
accreted by the new black hole. We have made the following assumptions
in our model: 
\begin{itemize} 
\item{The fraction of cold gas that forms stars over one 
dynamical time increases  with decreasing redshift.}
\item{A fraction $0.01/[1+(280/v_{\rm c})^2]$ of the cold gas 
in the merging galaxies is accreted by the black hole,
where $v_{\rm c}$ is the circular velocity of their combined
dark matter halo.}
\item{The accretion timescale of the gas 
scales with the dynamical time,
$t_{\rm acc}  = 2.5\times 10^7 [0.7+0.3(1+z)^3]^{-0.5} \haeyr $.}
\item{A fixed fraction of the accreted rest mass energy is radiated 
     away in the optical. The luminosity cannot exceed 
      the Eddington limit.}
\end{itemize}  
Our assumptions result in a strong decrease of the gas fraction 
in galaxies with redshift, from about 75 percent 
at $z=3$ to 10 percent at $z=0$.  Our model also fits the  strong 
decrease of the overall density of cool gas in the universe as 
inferred from the incidence rate of damped \haeop  absorbers  
(Fig 2.).  Because of  this change in gas fraction with redshift, 
the gas fraction in major mergers  that  produce bulges
is systematically  higher for  fainter  bulges, 
which form on average at higher redshift. This might 
explain the systematic differences in the the slope of the stellar 
density distribution in the cores of high-luminosity 
and low-luminosity ellipticals \cite{hae:mer99}.

\begin{figure}
\centering
\includegraphics[width=1.0\textwidth]{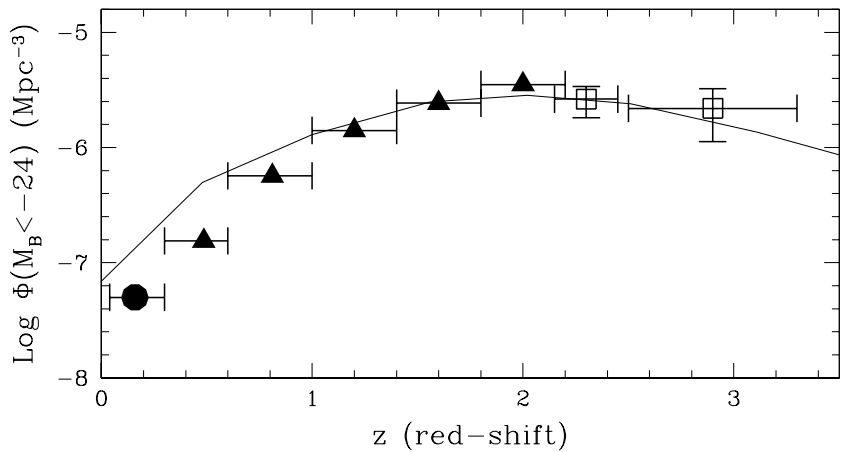}
\caption[]{
The evolution of the space density of quasars with $M_B < -24$. 
The data points are  a compilation by Grazian et al.  \cite{hae:gra99}} 
\end{figure}

\subsection{Remnant black holes in nearby galaxies}

Fig. 3 shows scatterplots of black hole mass versus
bulge luminosity drawn from  absolute magnitude-limited catalogues of 
bulges produced from our models.  The thick solid  line
shows the relation derived by Magorrian et al \cite{hae:mag98}
and the dashed  lines show the 1$\sigma$ scatter of their 
observational data around this relation. Both the slope and 
the scatter predicted by our models 
agree reasonably well with the observed relation. The normalization 
is set by the assumed fraction of the cool gas in merger    
remnants which is accreted onto the black hole. Up to one  
percent of the cold gas has to find its way into the central 
supermassive black hole to reproduce a present-day black hole to 
bulge mass ratio of 0.2 percent.  It is interesting to 
note that a central object with a  mass fraction of about   
two percent can stabilize a bar instability 
in a surrounding gas disc which may be responsible 
for driving the gas to the centre \cite{hae:shl90,hae:nsh96,hae:sel99}.  
One  observationally-testable prediction of our model is that
elliptical galaxies that formed recently should   harbour 
black holes with {\em smaller} masses than the spheroid 
population as a whole. This is illustrated in the second panel of
Fig. 3, where we show the relation between bulge luminosity 
and black hole mass for young  ellipticals.

\begin{figure}
\centering
\includegraphics[width=1.0\textwidth]{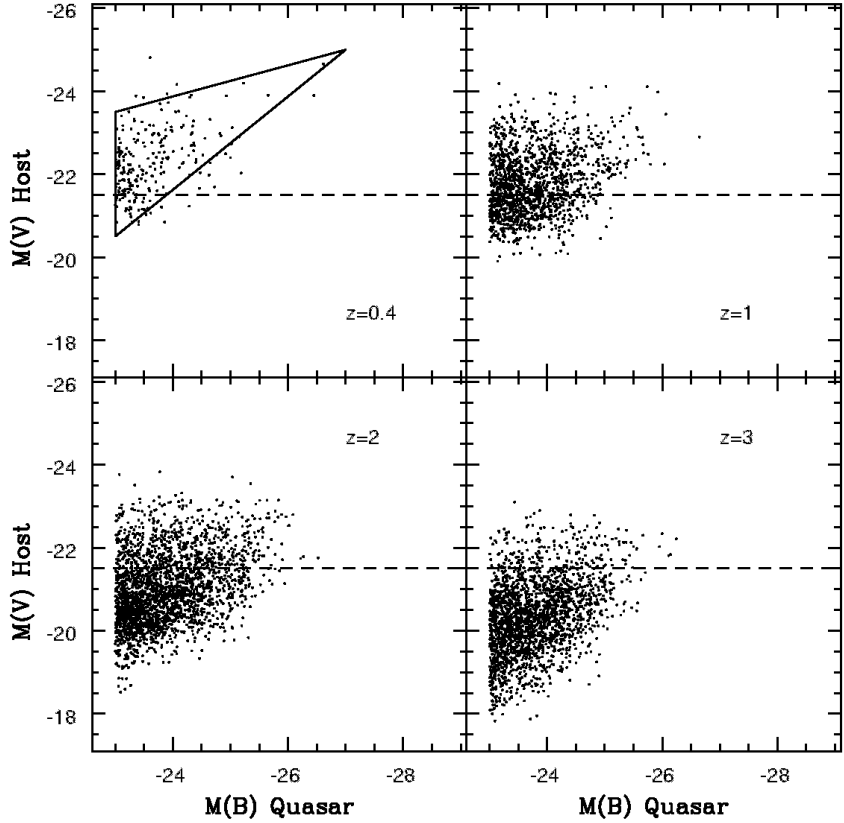}
\caption[]{Host galaxy versus quasar absolute
magnitudes at a series of redshifts. The dashed line shows the
present value of $L_*$ for galaxies. The triangular box in the top-left panel
shows the region spanned by the data set compiled by McLeod, Rieke \&
Storrie-Lombardi \cite{hae:mcl99}.}
\end{figure}

\subsection{Evolution of optically bright QSOs}

One of the striking features of the QSO population 
is their rapid evolution. The observed rapid decline of the 
space density at low redshift is not trivial to understand
\cite{hae:cav97}.  An important clue is probably  the 
similar rapid drop  of the overall amount of cool gas in the Universe 
inferred  from the rate of incidence   of damped 
\haeop systems.  The solid curve in Fig. 4 shows the model prediction 
for the evolution of the QSOs with $M_B < -24$ compared to
observational data compiled by Grazian et al. \cite{hae:gra99}.
The agreement is reasonably good.
In our model the strong decrease in quasar activity  
at low redshift results from a combination of three factors i) a
decrease in the  merging rates of intermediate mass galaxies at late
times,  ii) a decrease in the gas fraction of galaxies with 
decreasing redshift iii) the assumption  that black holes accrete gas 
more slowly at late times.

\subsection{QSO host galaxies}

In Fig. 5  we show scatterplots of host galaxy luminosity versus quasar
luminosity at a series of different redshifts. For reference, the
horizontal line in each plot shows present-day value of $L_*$ for galaxies.
At low redshift, quasars with magnitudes brighter than $M_B =
-23$ reside mostly in galaxies more luminous than
$L_*$.  Our results at low redshift agree remarkably well with a 
compilation of ground-based and HST observations of quasar hosts  by
Mcleod, Rieke \& Storrie-Lombardi \cite{hae:mcl99}.  
The triangle in Fig. 5 marks the region spanned by their 
observational data points.  At high
redshifts, our models predict that the quasars should be found in
progressively  {\em less luminous} host galaxies. This is not
surprising because in hierarchical models, the massive galaxies that
host luminous quasars at the present epoch are predicted to have
assembled recently \cite{hae:kch98}. 
The luminosities of quasars hosted by galaxies at different
epochs depends, however, on the redshift scaling of $t_{\rm acc}$.
Recently there have been a  number of detection of hosts of high 
redshift QSOs \cite{hae:are97,hae:rix99,hae:rid99}. Typically 
these seem to have $\sim L_*$ luminosity, suggesting that our assumed 
$t_{\rm acc}$ and its scaling with redshift  is indeed correct.

\section{Conclusions}

There is agreement to within a factor of a few between the black hole
mass density infered from black holes in nearby galaxies and that 
inferred from the radiation emitted by optically bright QSOs. 
This is  consistent with the possibility that optical bright 
quasars trace the assembly history of supermassive holes well,
but significant accretion in a different accretion mode with 
a different  redshift evolution is also viable. 

The large black hole to bulge mass ratio in nearby galaxies 
argues for a formation mechanism that avoids the intermediate 
step of a dense stellar cluster. A scenario in which supermassive 
black holes  are assembled by  mergers of smaller  black holes  
which formed well before the epoch  of optically bright QSOs 
would require high formation efficiencies (about 10 \%) in shallow 
potential wells. The most plausible mechanism by which the mass in a typical 
present-day supermassive black hole is assembled, 
is the collapse and accretion of cold gas plus some additional 
accretion of hot gas and merging of black holes at late times.

It is possible to built a unified model for the evolution of 
galaxies their central black holes and AGN activity by assuming that 
black holes  and bulges of galaxies form together during the frequent 
(major) mergers  predicted by hierarchical cosmogonies. 
Such a model can reproduce the observed rapid evolution of the space
density of bright QSO with redshift, the mass distribution of 
remnant black holes in nearby galaxies and the luminosity of 
QSO host galaxies.  

Interesting implications of our  model are the following. 
The typical duration of the optically bright QSO phase 
should be  $10^{7} \haeyr$. Young ellipticals should harbour black 
holes with smaller masses  than the spheroid  population as a whole.
QSO hosts are typically  brighter than $L_*$ at low 
redshift and should become  fainter with increasing redshift.  
Important for the rapid decline of the space density of bright QSOs
and the cosmological density of cold gas, is 
that the gas fraction in galaxies decreases
with decreasing  redshift. As a consequence, fainter 
ellipticals have formed in more  gas-rich mergers than 
bright ellipticals. Supermassive binaries and merging of
supermassive binaries should occur frequently in hierachical 
cosmogonies. The latter is  good news for space-borne gravitational 
wave experiments like LISA \cite{hae:hae94,hae:hae99}.

\vspace{0.5cm}

We acknowledge helpful discussions with Andrea Cattaneo, 
Stefano Cristiani, David Merrit, Prija Natarajan, Joel Primack, 
Martin Rees, Hans-Walter Rix and Simon White.

\clearpage
\addcontentsline{toc}{section}{Index}
\flushbottom
\printindex


\begin{thebibliography}{7}
%
\addcontentsline{toc}{section}{References}
\bibitem{hae:gen97} Genzel R., Eckart A., Ott T.,  Eisenhauer F., 1997, MNRAS,
201, 219 
\bibitem{hae:wat94} Watson W.D., Wallin B.K., 1994, ApJ, 432, L35
\bibitem{hae:miy95} Miyoshi M., Moran M., Hernstein J., Greenhill L.,
Nakai N., Diamond P., Inoue N., 1995, Nature, 373, 127  
\bibitem{hae:kor95} Kormendy J., Richstone D., 1995, ARAA, 33, 581 
\bibitem{hae:mag98} Magorrian J., et al., 1998, AJ, 115, 2285 
\bibitem{hae:vdm98} van der Marel R. P., 1998, in  Sanders D.B.,Barnes
J., eds, IAU Symposium 186 Kyoto 1997. Kluwer
\bibitem{hae:ric98} Richstone D., Ajhar, E.A., Bender, R., Bower, G.,
Dressler, A., Faber, S.M., Filippenko, A.V., Gebhardt, K. et al., 1998, 
Nat. Suppl., 395, 14
\bibitem{hae:chr99} Cattaneo,A., Haehnelt, M.G. \& Rees, M.J., 1999,
308, 77
\bibitem{hae:er88} Efstathiou G. P.,  Rees M. J., 1988, MNRAS, 230, 5p
\bibitem{hae:hr93} Haehnelt M.G.,  Rees M.J., 1993, MNRAS, 263, 168 
\bibitem{hae:kau96} Kauffmann, G., 1996, MNRAS, 281, 487 
\bibitem{hae:hai98} Haiman Z., Loeb A., 1998, ApJ, 503, 505
\bibitem{hae:kch98} Kauffmann, G. \& Charlot, S., 1998, MNRAS, 297, 23  
\bibitem{hae:bau98} Baugh, C.M., Cole, S., Frenk, C.G. \& Lacey, C.G., 1998,
ApJ, 498, 504
\bibitem{hae:spf99} Somerville, R.S., Primack, J.R. \& Faber, S.M.,
1999, submitted, astro-ph/9806228
\bibitem{hae:kha99} Kauffmann G., Haehnelt M., 1999, MNRAS, in press
\bibitem{hae:sal99} Salucci P., Szuszkievicz E., Monaco, P., Danese, L., 1999,
MNRAS, in press, astro-ph/9811102
\bibitem{hae:fhp98} Fugukita M., Hogan C.J.,  Peebles P.J.E., 1998,
ApJ, 503, 518 
\bibitem{hae:sol82} Soltan A., 1982, MNRAS, 200, 115.
\bibitem{hae:cht92} Chokshi A.,  Turner E. L., 1992, MNRAS, 259, 421 
\bibitem{hae:dim97} Di Matteo T.,  Fabian A. C., 1997, MNRAS, 286, 393
\bibitem{hae:alm99} Almaini O., Lawrence A., Boyle B.J., 1999, MNRAS,
305, L59
\bibitem{hae:pug96} Puget J.L., Abergel A., Bernard J.P., Boulanger F.,
Burton W.B.,  Desert F.X., Hartmann D., 1996, A\&A, 308, L5
\bibitem{hae:phi97} Phinney E. S., 1997, talk presented at the IAU Symposium
186 Kyoto 1997 
\bibitem{hae:hnr98} Haehnelt M., Natarajan P.,  Rees M.J., 1998,
MNRAS, 300, 817
\bibitem{hae:ree84} Rees M. J., 1984, ARAA, 22, 471
\bibitem{hae:qui90} Quinlan G.D., Shapiro S.L., 1990, ApJ, 356, 483
\bibitem{hae:beg80} Begelman M.C., Blandford R.D., Rees  M.J., 1980, Nature, 287, 307  
\bibitem{hae:sto96} Storrie-Lombardi, L.J., MacMahon, R.G. \& Irwin, M.J., 1996, MNRAS, 283, L79
\bibitem{hae:mer99} Meritt D., 1999, in:''Galaxy Dynamics: From the Early 
Universe to the Present'', Paris 1999, eds. F. Combes, G. Mamon, 
V. Charmandaris
\bibitem{hae:shl90} Shlosman I., Begelman M.C., Julian F., 1990, Nat., 345, 679
\bibitem{hae:nsh96} Norman C., Sellwood J.A., Hasan H., 1996, ApJ,
462, 114
\bibitem{hae:sel99} Sellwood J.A., Moore E.M., 1999, ApJ, 510, 125

\bibitem{hae:cav97} Cavaliere A.,  Vittorini V., 1997,  in M\"uller V. et al.,
1997,  Proc. 12th Potsdam cosmology workshop., astro-ph/9712295 
\bibitem{hae:gra99} Grazian, A., Cristiani, S., D'Odorico, V., Omizzolo,
A., \& Pizella, A., 2000, preprint
\bibitem{hae:mcl99} McLeod, K.K., Rieke, G.H. \& Storrie-Lombardi, L.J.,1999, ApJ, 511, L67
\bibitem{hae:are97} Aretxaga I., Terlevich R.J., Boyle B.J., 1997,
MNRAS, 
\bibitem{hae:rix99} Rix H.-W., Falco E., Impey C., Kochanek C., Lehar
J., McLeod B., Munoz J., Peng C., 1999, in ``Gravitational Lensing: Recent
Progress and Future Goals'', eds. T. Brained amd C. Kochanek, astroph/9910190
\bibitem{hae:rid99} Ridgway S., Heckman T., Calzetti D., Lehnert M., 
1999, in;'' Lifecycles of Radio Galaxies'', eds. J. Biretta et al, 
New Astronomy Reviews, astroph/9911049 
\bibitem{hae:hae94} Haehnelt M.G., 1994,  MNRAS, 269, 199 
\bibitem{hae:hae99} Haehnelt M.G., 1999 in:``The second international LISA
symposium'',  Pasadena 1998, Ed. Folkner W., AIP, p. 45  

 


\end{thebibliography}
\end{document}